\documentclass[9pt,conference,onecolumn]{IEEEtran}
\usepackage[utf8]{inputenc}
\usepackage[english]{babel}
\usepackage{graphicx}
\usepackage{amsmath}
\usepackage{tikz}
\usepackage{array}
\usepackage{booktabs}
\usepackage{float}
\usepackage{amssymb} %\mathcal{}
\usepackage{mathrsfs} %\mathscr{}
\usepackage{xargs}
\usepackage{dblfloatfix}
\usepackage[caption=false,font=footnotesize]{subfig}
\usepackage{siunitx} %\SI
\usepackage{tikzscale}

\usepackage{mathtools} %for stacking equal sign
\usepackage{commath} %for abs value and norm
\newcommand{\lefto}{\mathopen{}\left}
\newcommand{\righto}{\mathopen{}\right}

\definecolor{darkblue}{RGB}{0, 5, 150}

\providecommand{\vectornorm}[1]{\left\lVert#1\right\rVert}

\allowdisplaybreaks

\title{Massive MIMO for Ultra-reliable Communications with Constellations for Dual Coherent-noncoherent Detection}
\author{\IEEEauthorblockN{Alexandru-Sabin Bana, Marko Angjelichinoski, Elisabeth de Carvalho, Petar Popovski}
\IEEEauthorblockA{Department of Electronic Systems, Aalborg University, Denmark}}

\begin{document}
%\listoftodos % todos
\maketitle
\begin{abstract}
The stringent requirements of ultra-reliable low-latency communications (URLLC) require rethinking of the physical layer transmission techniques. 
Massive antenna arrays are seen as an enabler of the emerging $5^\text{th}$ generation systems, due to increases in spectral efficiency and degrees of freedom for transmissions, which can greatly improve reliability under demanding latency requirements. 
Massive array coherent processing relies on accurate channel state information (CSI) in order to achieve high reliability. 
In this paper, we investigate the impact of imperfect CSI in a single-input multiple-output (SIMO) system on the coherent receiver. 
An amplitude-phase keying (APK) symbol constellation is proposed, where each two symmetric symbols reside on distinct power levels.
The symbols are demodulated using a dual-stage non-coherent and coherent detection strategy, in order to improve symbol reliability. 
By means of analysis and simulation, we find an adequate scaling of the constellation and show that for high signal-to-noise ratio (SNR) and inaccurate CSI regime, the proposed scheme enhances receiver performance.
\end{abstract}

\section{Introduction}
Fifth generation (5G) communication systems promise to serve not only high-throughput applications, known as extended mobile broadband (eMBB), but also two main instances of machine-type communication (MTC): massive MTC (mMTC) and ultra-reliable low-latency communications (URLLC) \cite{towards_massiveURLLC_short_pkts}. 
Being the first communication system to specifically address the requirements of mMTC and URLLC and aiming to integrate the aforementioned three services, it is evident that the physical layer must serve as a common ground for these services to coexist \cite{waveform_numerology_5g_services}. 
However, the increased number of devices of mMTC, the stringent latency-reliability requirement of URLLC and the short packet nature of the MTC transmissions in general require reevaluating physical layer design and processing \cite{pkt_structure_urllc}. 
The importance of exploiting various diversity resources in order to increase reliability without increasing latency is discussed in \cite{URLLC_principles_magazine}, where also several design principles are presented. Massive multiple-input multiple-output (MIMO) is seen as a prominent enabler for URLLC \cite{feasibility_large_arrays_urllc}, \cite{urllc_mmwave_massivemimo}, \cite{five_disruptive_tech_5g}, as well as for mMTC \cite{massivemimo_massive_connect} and eMBB \cite{massiveMIMO_maximal_SE}.
Multiple antennas at both receiver and transmitter enable exploiting favorable propagation conditions \cite{favorable_propag}, meaning having the ability to use the additional degrees of freedom (DoFs) to spatially separate users.
Furthermore, an additional benefit inherited from the massive arrays is the hardening of the channel and noise \cite{massive_mimo_next_generation}, \cite{massive_mimo_properties_measured_channels}. 
However, the convenient properties of massive arrays are not to be taken for granted, especially if coherent reception takes place and the CSI is either outdated or imprecise \cite{massive_mimo_next_generation}.

These imperfections of the CSI are the core motivation for our interest in investigating robust receiver processing techniques. 
In non-coherent energy detection (ED) receivers, the information is contained in the symbol energy levels, while in coherent phase-shift keying (PSK), information is contained in the symbol phase.
Drawing inspiration from non-coherent (ED) receivers with massive arrays \cite{noncoherent_design_performance} and earlier works on constellation designs \cite{digital_amplitude_phase},\cite{variable_rate_qam}, we propose a \emph{dual-stage}, \emph{sequential} non-coherent and coherent detection strategy, where the receiver firstly performs non-coherent detection of the energy level to narrow down the possible received symbols, and then uses the CSI to coherently detect the exact symbol. 
The non-coherent detection benefits from the massive MIMO effect, as it does not rely on exact estimation of the channel coefficients for each receiver antenna, requiring only the estimate of the channel energy over the array. Due to the very large number of antennas, the needed channel energy tends to be very robust to channel small scale variations. 
The key improvement of the proposed approach is that in the first stage, the uncertainty in the received symbol is reduced by using the channel energy, such that in the second stage, even if the CSI is imprecise, the chance of correctly detecting the received symbol is improved, since there are fewer possible symbols.
The proposed scheme is evaluated through analysis and simulation and is shown to improve receiver performance for high SNR scenarios where CSI is inaccurate. 

\section{System model}
This paper considers an uplink (UL) URLLC scenario, where a single-antenna device transmits a short message to a massive array receiver, which is to be decoded with very high reliability. It is assumed that the transmission has been previously scheduled and that it is subject to fading and noise, such that the received signal at each receiver antenna for a transmitted symbol $x$ is expressed as
\begin{align}
    y_i = h_i x + n_i.
\end{align}
The vectors $\textbf{y},\textbf{h}$ and $\textbf{n}$ denote the received signal, the channel coefficients and the noise, respectively, for each antenna $i\in\{1,\dots,M\}$. Both the channel coefficients and the noise are assumed to be circularly symmetric i.i.d. complex Gaussian random variables.

\begin{figure}[t]
    \centering
    \includegraphics[width=0.7\columnwidth]{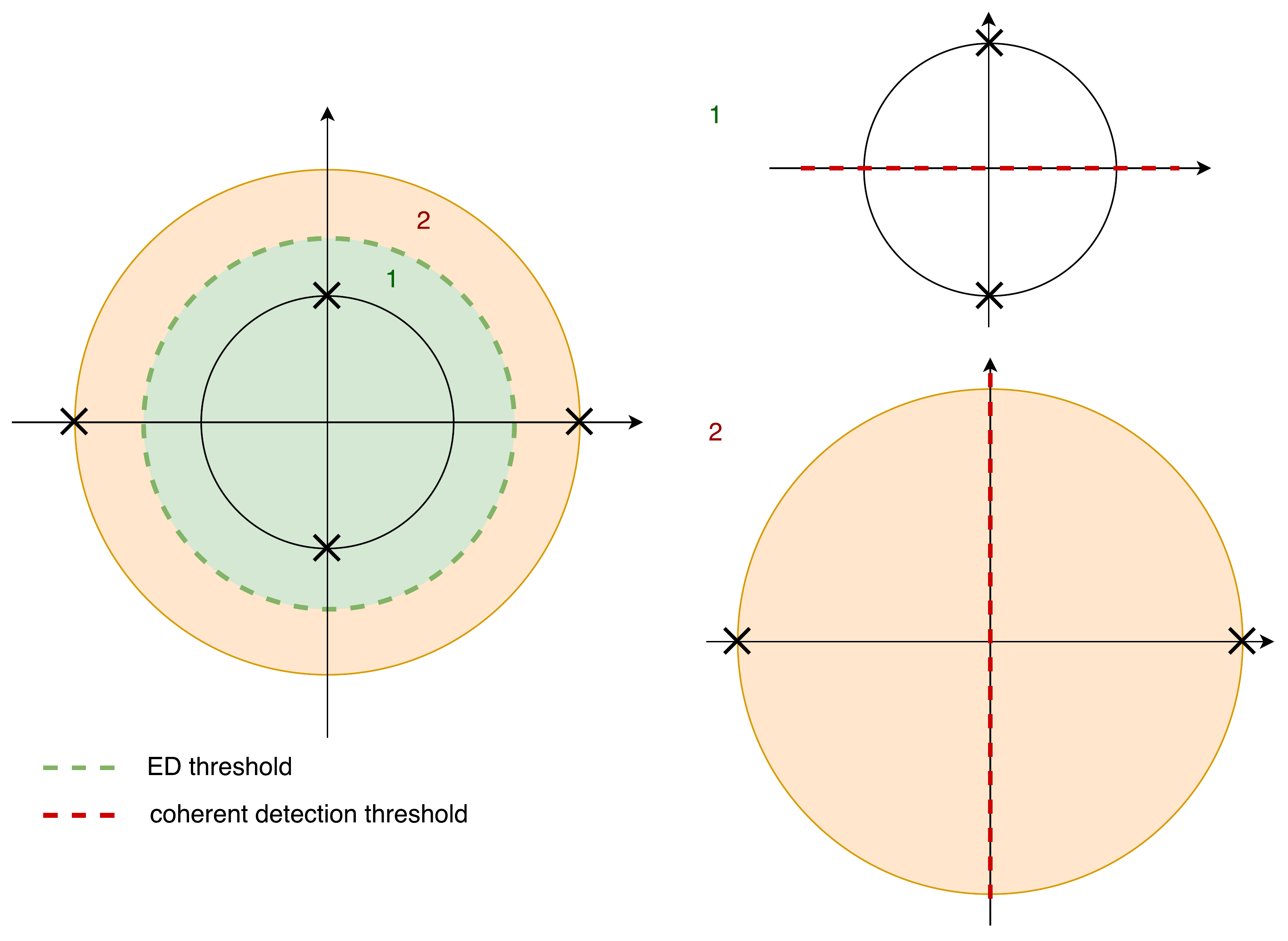}
    \caption{Proposed 4-APK constellation, where each two symmetric symbols reside on distinct power levels, and representation of the dual-stage sequential non-coherent and coherent detection thresholds.}
    \label{fig:sys_model}
\end{figure}

We use 4-ary modulation candidates for the set of possible transmitted symbols $x$, such as phase-shift keying (more precisely, 4-PSK) and amplitude-phase keying (APK), as described in \cite{digital_amplitude_phase} and shown in \ref{fig:sys_model}. 
For the case of 4-PSK, the symbol alphabet is $\{\pm 1,\pm i\}$, whereas when using APK modulation, one can use different symbol amplitudes, such as $\{\pm \sqrt{(1+s)},\pm i\sqrt{(1-s)}\}$, where we denote $s$ as the scaling factor, such that the average power of the alphabet remains the same as for the 4-PSK.
Furthermore, we assume that the symbols are equiprobable, such that the power constraint per transmission is satisfied with a very high probability.

The coherent receiver, which is the benchmark, performs maximum ratio combining (MRC) of the signals received over the large array; the resulting signal can be expressed as
\begin{align}
    y_\text{MRC}= \dfrac{\sum_{i=1}^M \hat{h}_i^* y_i}{\sum_{i=1}^M \abs{\hat{h}_i}^2} \label{eq:y_mrc}
\end{align}
where $\hat{h}$ is the pilot-based estimate of the channel coefficients, which cannot be considered ideal, i.e., $\hat{h}\neq h$ since several factors such as high mobility, limited training resources in stringent latency conditions, and neighboring cell interference can lead to having either outdated or contaminated CSI. 
Therefore, we model the CSI impairments similarly to earlier works \cite{large_system_analysis_miso}, as
\begin{align}
    \hat{h} = \rho h + \sqrt{1-\rho^2} \gamma \label{eq:csi_est}
\end{align}
where $\gamma$ is a circularly symmetric i.i.d. complex Gaussian random variable, with zero mean and the same variance as $h$.
For $\rho\approx 1$, the CSI is close to ideal.

The proposed receiver consists of two stages performing sequential detection of the symbols, as illustrated in Fig.~\ref{fig:sys_model}. 
For the first stage, energy detection is performed in order to determine to which level the received signal belongs to. 
Secondly, after narrowing down the possible symbols to half, coherent maximum-likelihood (ML) detection is performed on the maximum-ratio combined signal. Instantaneous channel energy based ED (I-ED) \cite[Section IV]{noncoherent_design_performance} is chosen for the first stage of the detection since it only requires estimates of the sum of squared amplitudes of the instantaneous channel coefficients, and the noise variance. This channel energy is more stable because of the large number of antennas, making I-ED suitable for imprecise CSI scenarios. As a result, I-ED exploits the massive MIMO effect when estimating the channel energy and the fact that the noise is independent and identically distributed across receive antennas, therefore achieving noise hardening.

\section{Analysis}
We evaluate analytically the performance of the well-known coherent MRC receiver, the non-coherent I-ED receiver \cite{noncoherent_design_performance} and the proposed dual-stage receiver under imperfect channel conditions.
\subsection{Coherent 4-PSK MRC receiver}
\begin{figure}[t]
    \centering
    \includegraphics[width=0.7\columnwidth]{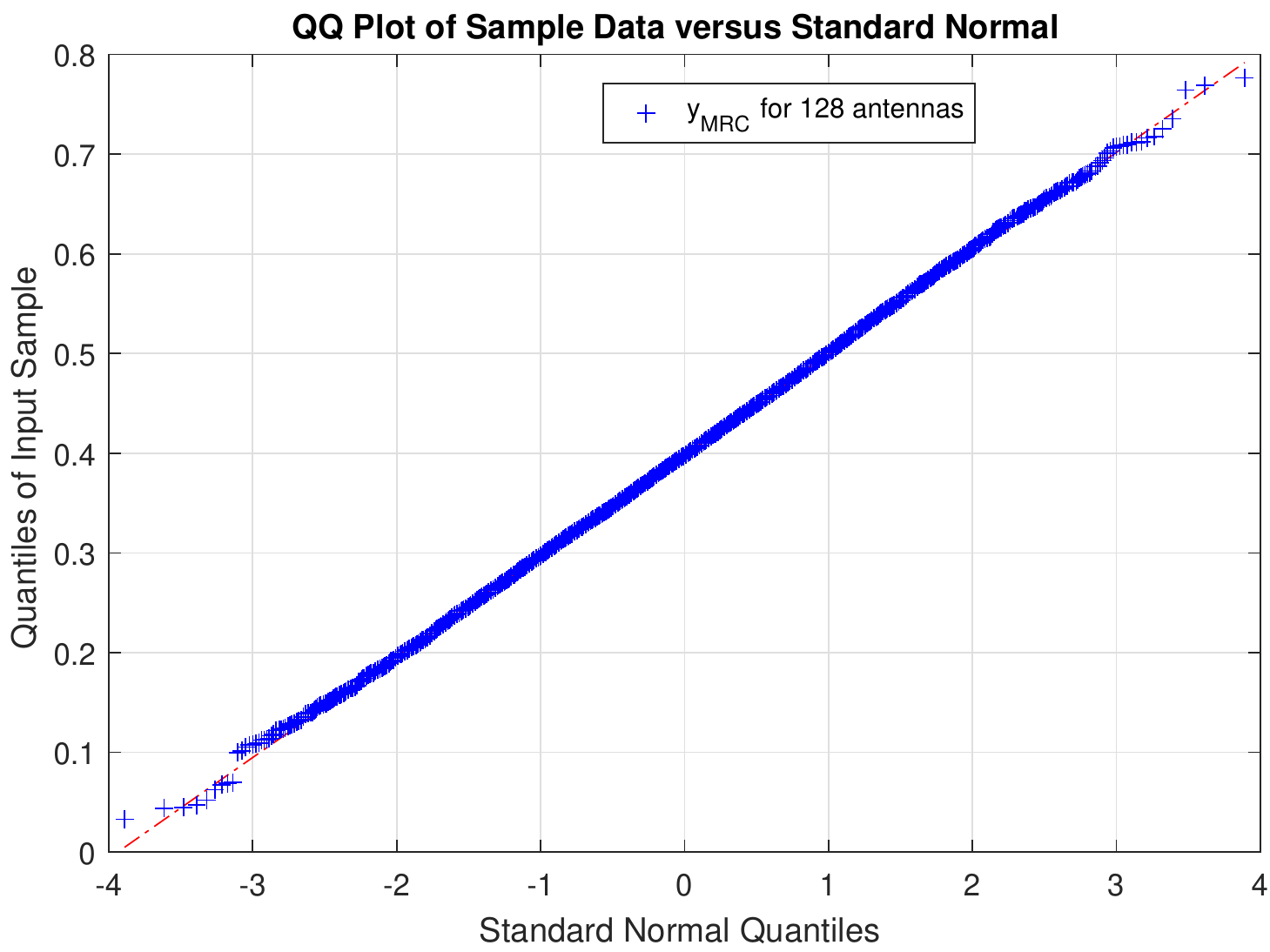}
    \caption{Plot of quantiles of $y_\text{MRC}$ versus standard normal quantiles. It can be seen that the distribution of $y_\text{MRC}$ can be well approximated by a standard normal distribution.}
    \label{fig:qqplot}
\end{figure}

Given the received signal expression for the MRC case in \eqref{eq:y_mrc}, and the CSI impairments model in \eqref{eq:csi_est}, the exact distribution of $y_\text{MRC}$ cannot be easily deduced. However, by using the central limit theorem, it is observed that the distributions of the real and imaginary part of $y_\text{MRC}$ can be well approximated by a normal distribution. This is shown in Fig.~\ref{fig:qqplot}.

In order to express the symbol error rate (SER), it is further needed to determine the mean and variance of $y_\text{MRC}$. These can be obtained either through Monte-Carlo simulations or by using Rao's procedure for determining the mean and variance of ratios of random variables \cite{frishman1971arithmetic}.

Due to the symmetry of the constellation, the mean and variance are the same for both real and imaginary domain, meaning that the SER can be expressed by using the tail distribution function of the standard normal distribution $\left(Q(x)\right)$ as
\begin{IEEEeqnarray}{l}
    \text{SER}^\text{4-PSK} =
    \approx 2Q\lefto(\dfrac{\mu_{y_\text{MRC}}}{\sigma_{y_\text{MRC}}}\righto). \label{eq:p_e_coh}
\end{IEEEeqnarray}
% \begin{}[t]
%     \centering
%     \includegraphics[width=1\columnwidth]{figures/mean_var_mrc_approx}
%     \caption{Comparison between sample mean and variance and their approximations based on \eqref{eq:mean_approx}-\eqref{eq:var_approx}. It can be noticed that the mean is well approximated, whereas the variance is sufficiently accurate only for low values $\rho$.}
%     \label{fig:mean_var_approx}
% \end{figure}

\begin{figure*}[t]
\centering
\subfloat[$\text{SNR}=\SI{-2}{dB}$ ]{\includegraphics[width=0.5\columnwidth]{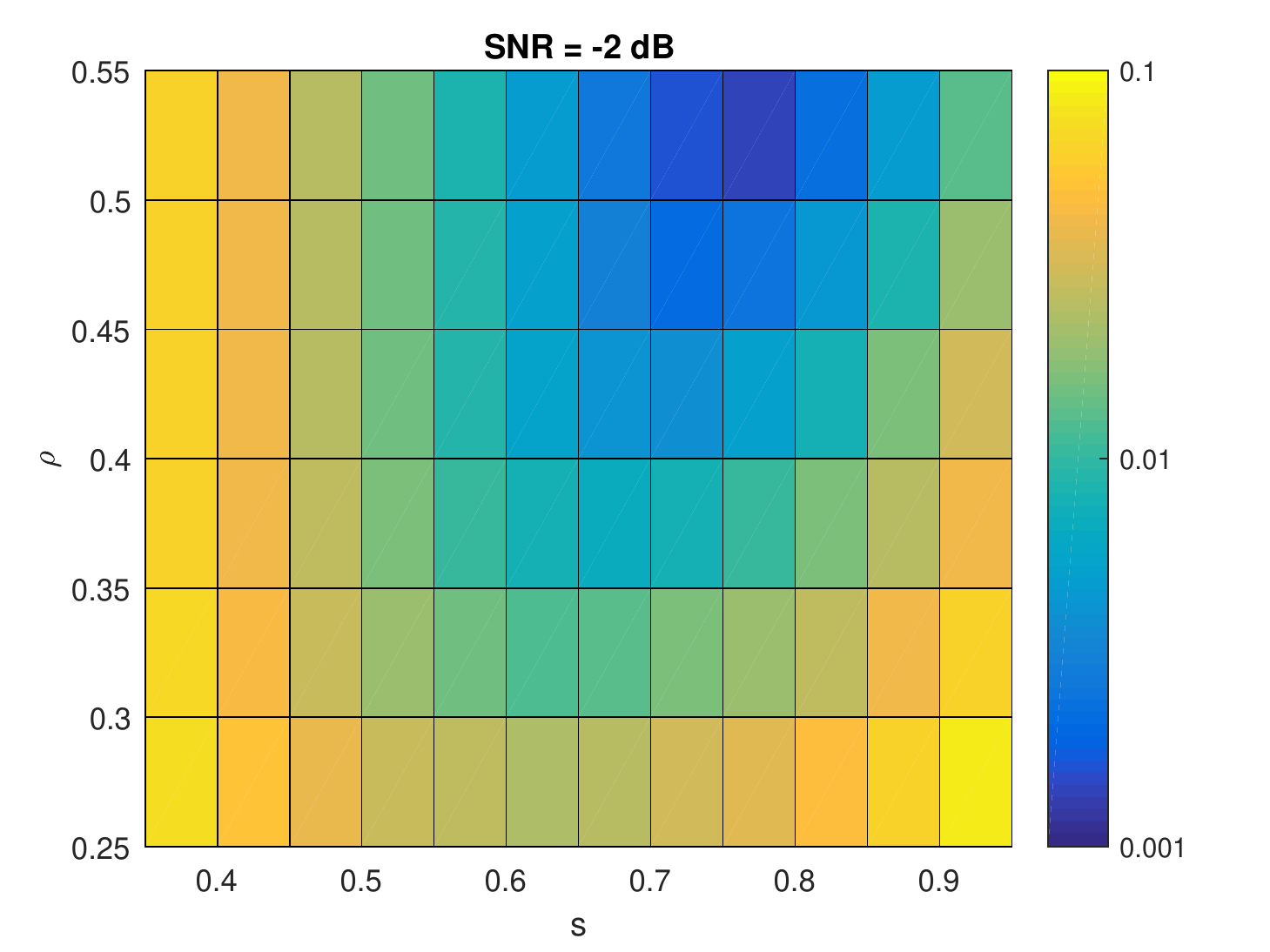}\label{fig:scaling-2}}
\hfil
\subfloat[$\text{SNR}=\SI{2}{dB}$]{\includegraphics[width=0.5\columnwidth]{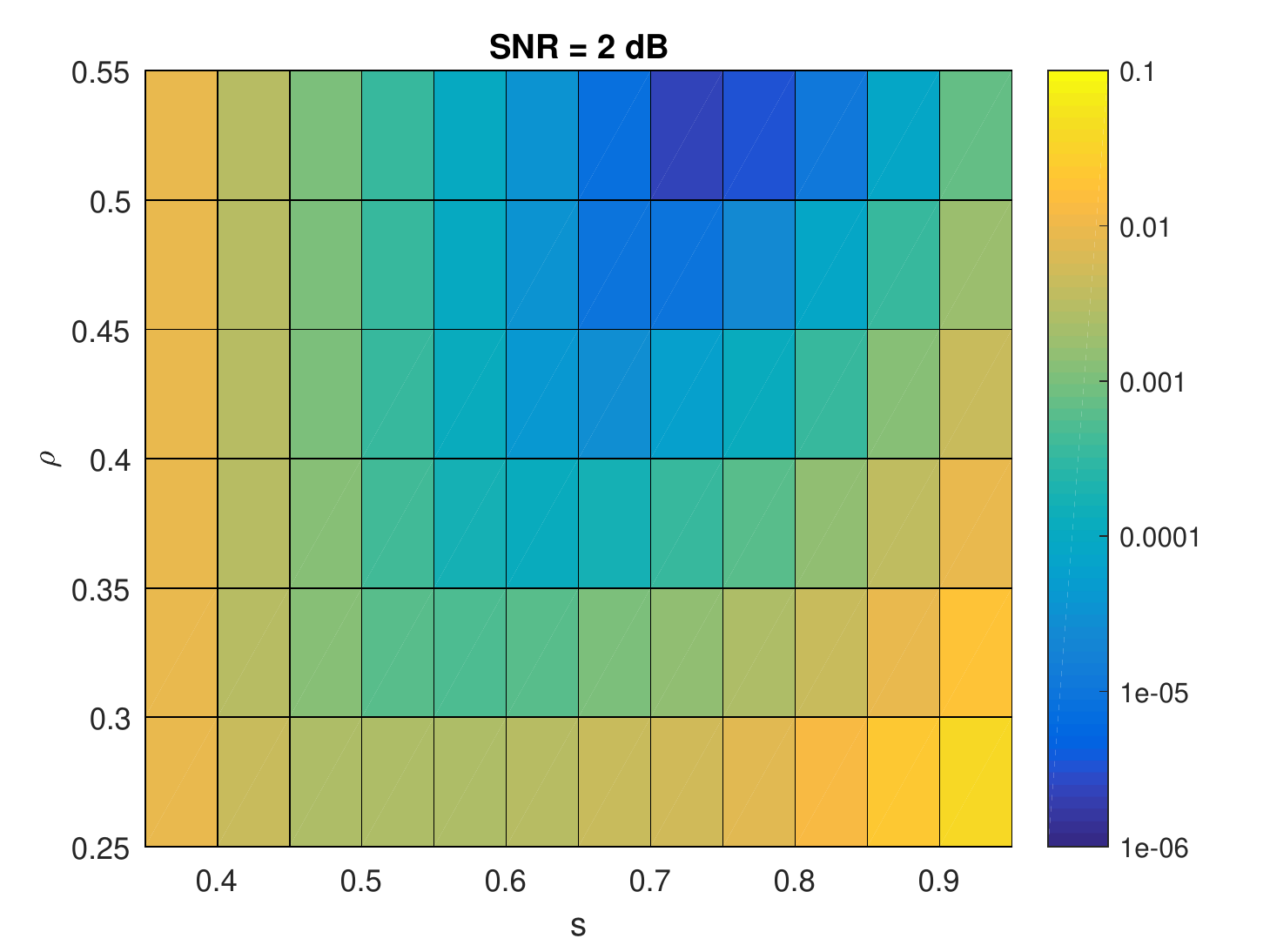}\label{fig:scaling2}}
\hfil
\subfloat[$\text{SNR}=\SI{0}{dB}$]{\includegraphics[width=0.5\columnwidth]{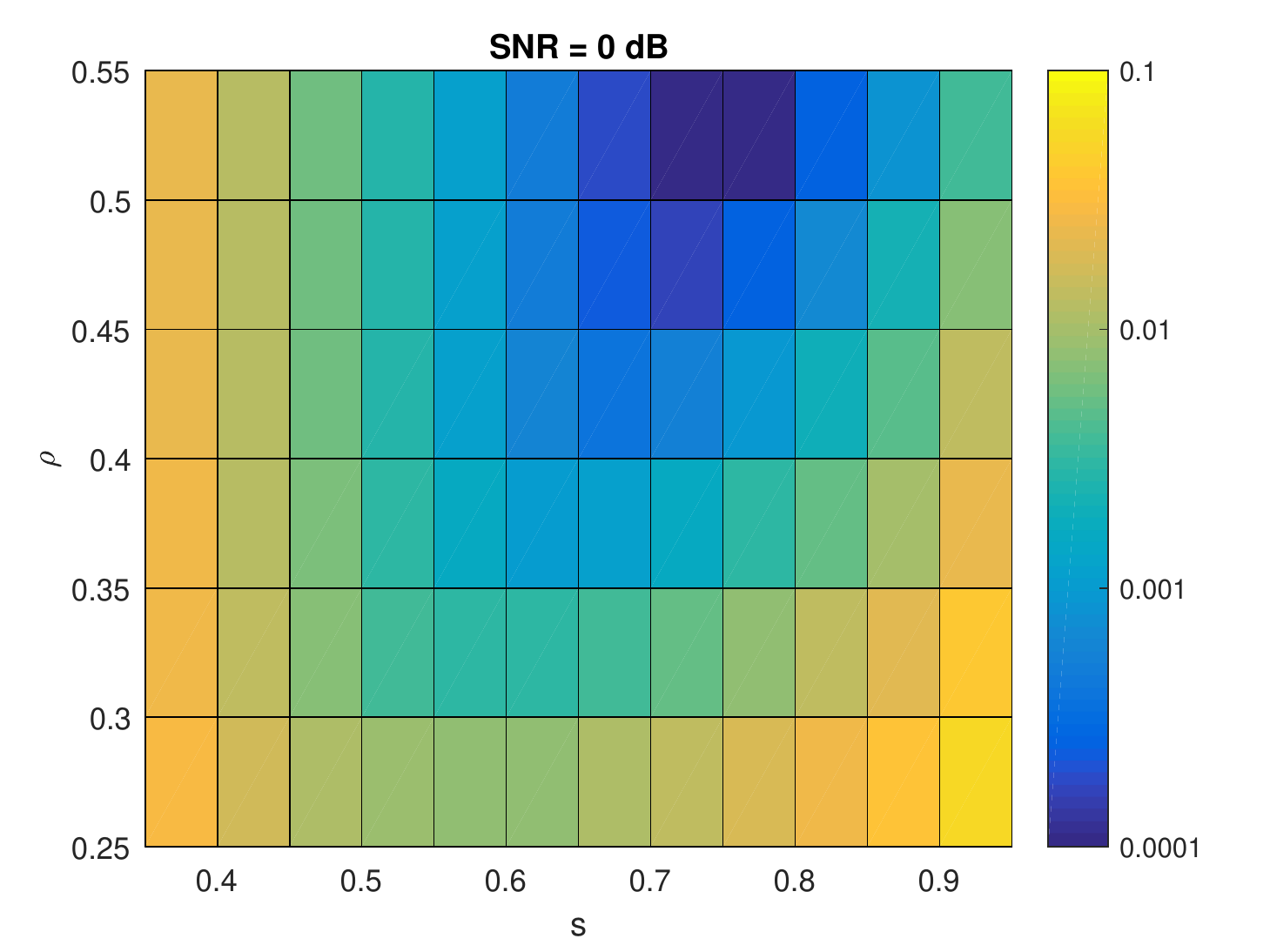}\label{fig:scaling0}}
\hfil
\subfloat[$\text{SNR}=\SI{4}{dB}$]{\includegraphics[width=0.5\columnwidth]{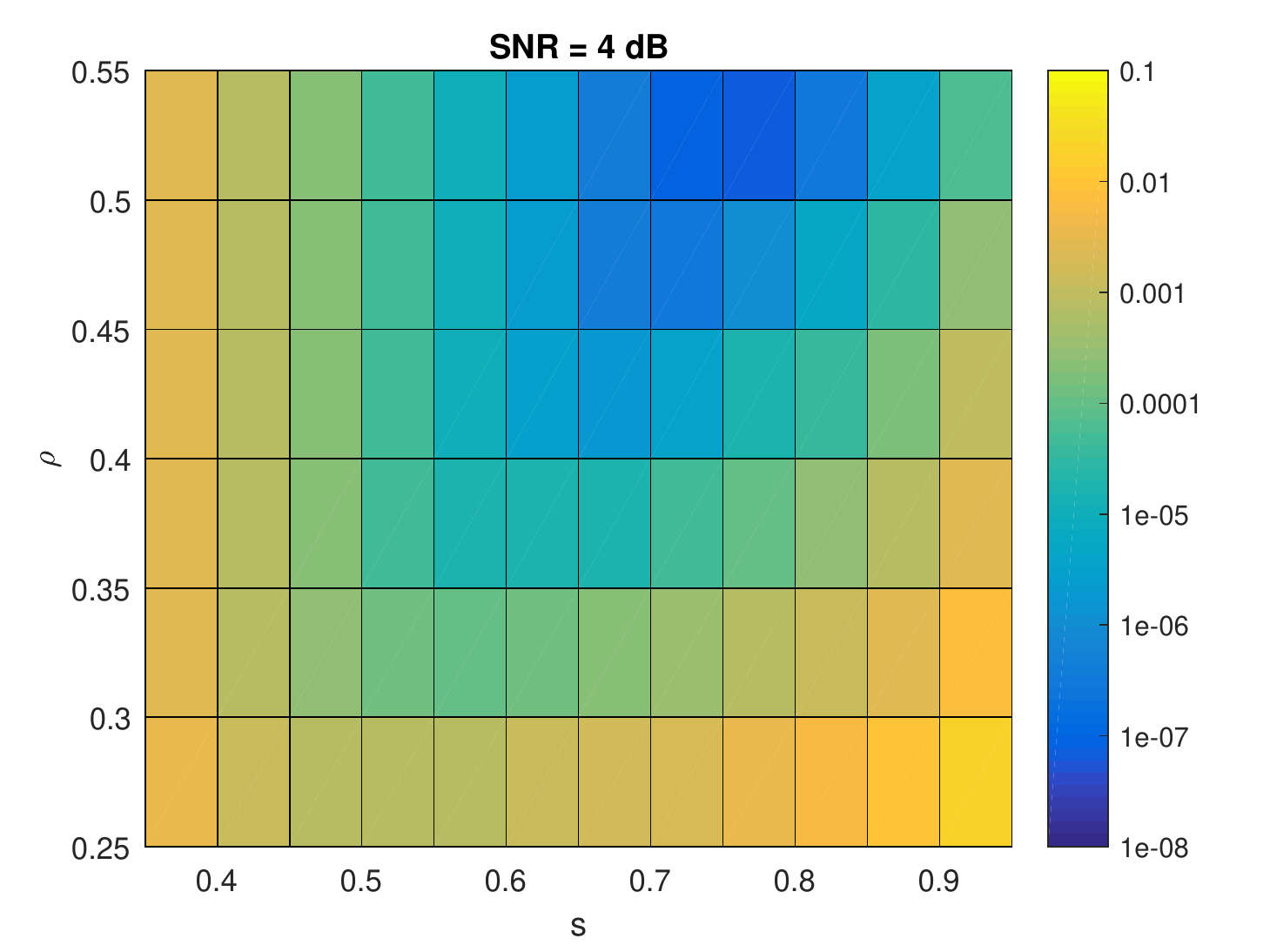}\label{fig:scaling4}}
\caption{Dependency of the symbol error rate on the choice of the constellation scaling coefficient.}
\label{results2}
\end{figure*}

\subsection{Non-coherent I-ED receiver}
The non-coherent instantaneous energy detection (I-ED) receiver relies on estimates of the instantaneous channel energy $(\varsigma_h)$ in order to detect the energy of the transmitted symbols, which is expressed in \cite{noncoherent_design_performance} as
\begin{IEEEeqnarray}{ll}
z = \underbrace{\dfrac{1}{M} \vectornorm{\mathbf{h}}_2^2}_{\varsigma_h} x^2 + \underbrace{\dfrac{1}{M} \vectornorm{\mathbf{n}}_2^2}_{\varsigma_n} + \underbrace{2\dfrac{1}{M} \Re[\mathbf{h}^H\mathbf{n}]}_{w}x.
\end{IEEEeqnarray}

It has been shown in \cite{noncoherent_design_performance} that as $M\rightarrow \infty$, $\varsigma_n$ approaches the noise variance $\sigma_n^2$ and that the cross term $w$ tends to zero, due to the law of large numbers. Therefore, noise hardening occurs, meaning that the noise contribution becomes more deterministic, and can therefore be suppressed if $\sigma_n^2$ is estimated.

The received energy $z$ follows a non-central Chi-square distribution. However, according to the central limit theorem, when $M$ is large, $z$ is well approximated by a non-central Gaussian variable with the same mean and variance as the non-central Chi-square distribution \cite{noncoherent_design_performance}, which are equal to
\begin{IEEEeqnarray}{ll}
\mu_z (\varsigma_h,\epsilon_p) &= \mu_{z,p} = \varsigma_h \epsilon_p + \sigma_n^2 ; \\
\sigma_z^2(\varsigma_h,\epsilon_p) &= \sigma_{z,p}^2 = \dfrac{\sigma_n^2}{M}(2\varsigma_h\epsilon_p + \sigma_n^2)
\end{IEEEeqnarray}
where $\epsilon_p$ is the power of a symbol and for 2-PAM, $p\in\{0,1\}$.
For details on how the detection thresholds are computed, the reader can refer to \cite[Section IV.B]{noncoherent_design_performance}. Under the Gaussian approximation assumption, the SER for a 2-PAM constellation can be expressed as
\begin{IEEEeqnarray}{ll}
	\text{SER}^{\text{I-ED}}(\varsigma_h)\approx \dfrac{1}{2} \lefto[ Q\lefto( \dfrac{\Delta_p(\varsigma_h) - \mu_{z,p}}{\sigma_{z,p}}\righto) + Q\lefto( \dfrac{\mu_{z,p+1} - \Delta_p(\varsigma_h)}{\sigma_{z,p+1}}\righto)\righto].\IEEEeqnarraynumspace \label{eq:p_e_i-ed}
\end{IEEEeqnarray}

\subsection{Dual-stage sequential receiver}
The overall SER of the proposed receiver can be expressed as
\begin{IEEEeqnarray}{l}
\text{SER}^\text{dual} = \text{SER}^{\text{I-ED}} + (1-\text{SER}^{\text{I-ED}})  \text{SER}^{\text{coh}}.
\end{IEEEeqnarray}
The symbol error probability in the first stage $( \text{SER}^{\text{I-ED}})$ of the sequential detection is characterized by \eqref{eq:p_e_i-ed}, whereas the SER in the coherent detection stage $(\text{SER}^{\text{coh}})$ can be expressed as 
\begin{IEEEeqnarray}{l}
\text{SER}^{\text{coh}} = \dfrac{1}{2}\lefto(\text{SER}^{\text{re}} + \text{SER}^{\text{im}} \righto).\label{eq:pe_coh_dual}
\end{IEEEeqnarray}
The $1/2$ coefficient in \eqref{eq:pe_coh_dual} stems from the probability of having to decide either from two real- or imaginary-valued symbols (see Fig.~\ref{fig:sys_model}). Further, based on the Gaussian approximation of $y_\text{MRC}$, the two probabilities inside the bracket can be expressed as:
\begin{IEEEeqnarray}{ll}
	\text{SER}^{\text{re}} &= \dfrac{1}{2}\lefto(\Pr\lefto[\Re[y_\text{MRC}]^+ <0 \righto] + \Pr\lefto[\Re[y_\text{MRC}]^- >0 \righto]\righto) \\
	&= \dfrac{1}{2} \lefto[ Q\lefto( \dfrac{\mu_\text{re}(\sqrt{1+s})}{\sigma_\text{re}(\sqrt{1+s})} \righto) + Q\lefto(-\dfrac{\mu_\text{re}(-\sqrt{1+s})}{\sigma_\text{re}(-\sqrt{1+s})} \righto)\righto], \IEEEeqnarraynumspace \label{eq:p_e_re}
\end{IEEEeqnarray}
where $\Re[y_\text{MRC}]^+$ and $\Re[y_\text{MRC}]^-$ denote the real value of the received signal if the corresponding real-valued positive or negative symbol was sent, respectively. Similarly, $\text{SER}^\text{im}$ can be expressed as 
\begin{IEEEeqnarray}{ll}
	\text{SER}^\text{im} &= \dfrac{1}{2} \lefto(\Pr\lefto[\Im[y_\text{MRC}]^+ <0 \righto] + \Pr\lefto[\Im[y_\text{MRC}]^- >0 \righto]\righto) \\
	&=\dfrac{1}{2}\lefto[ Q\lefto( \dfrac{\mu_\text{im}(\sqrt{1-s})}{\sigma_\text{im}(\sqrt{1-s})} \righto) + Q\lefto(-\dfrac{\mu_\text{im}(-\sqrt{1-s})}{\sigma_\text{im}(-\sqrt{1-s})} \righto) \righto]. \IEEEeqnarraynumspace
\end{IEEEeqnarray}
One can note that the two terms $\text{SER}^\text{re}$ and $\text{SER}^\text{im}$ would be identical if the power of the constellation symbols were not scaled by $s$, therefore equation \eqref{eq:pe_coh_dual} would simply become the same as for the 4-PSK case \eqref{eq:p_e_coh}. Due to the dependency of the SER on the scaling of the constellation, we study the trade-off of choosing $s$ in the following section.

\section{Constellation scaling and results}
The constellation scaling coefficient $s$ has distinct impact on each of the two stages of the sequential decoder. On one hand, increasing the scaling offers a larger distance between the power levels for the first (non-coherent) stage, thereby improving the reliability of detection. However, for the second (coherent) stage, increasing the scaling offers higher distance only between two of the constellation points, the ones lying on the outer power level, whereas the symbols that were scaled down will be more likely not to be decoded properly. Essentially, the more scaling is applied, the worse the coherent detection becomes due to straying away from the optimal 4-PSK.

This trade-off induced by the choice of the scaling coefficient is illustrated in Figures \ref{fig:scaling-2}-\ref{fig:scaling4}, using the analysis results for the SER from the previous section. It can be noticed that the optimal constellation scaling is centered around values of 0.7, and it does not seem to depend on the CSI correlation $\rho$.
For two of the SNR and CSI accuracy regimes illustrated, a constellation scaling of 0.7 offers symbol error rates below $10^{-5}$. %We should highlight that SER for the noncoherent stage in the figure is based on a 2-level energy detection, which is why it is plotted as the baseline for the dual-stage.

% \begin{figure}[t]
%     \centering
%     \includegraphics[width=1\columnwidth]{figures/choice_scaling_v3}
%     \caption{Dependency of the symbol error rate on the choice of the constellation scaling coefficient for the noncoherent stage (solid lines) and the dual-stage receiver (dashed). Distinct SNRs are represented in different colors, and different CSI correlation coefficient $\rho$ are shown with different markers.}%\textcolor{blue}{The function is imagesc(A), A is 2D array (i.e., over $s$ and $\rho$). Make sure to have high resolution. Plot the SER for 6 different SNR values. You can even put larger range for s.}
%     \label{fig:scaling}
% \end{figure}

\begin{figure}[t]
    \centering
    \includegraphics[width=1\columnwidth]{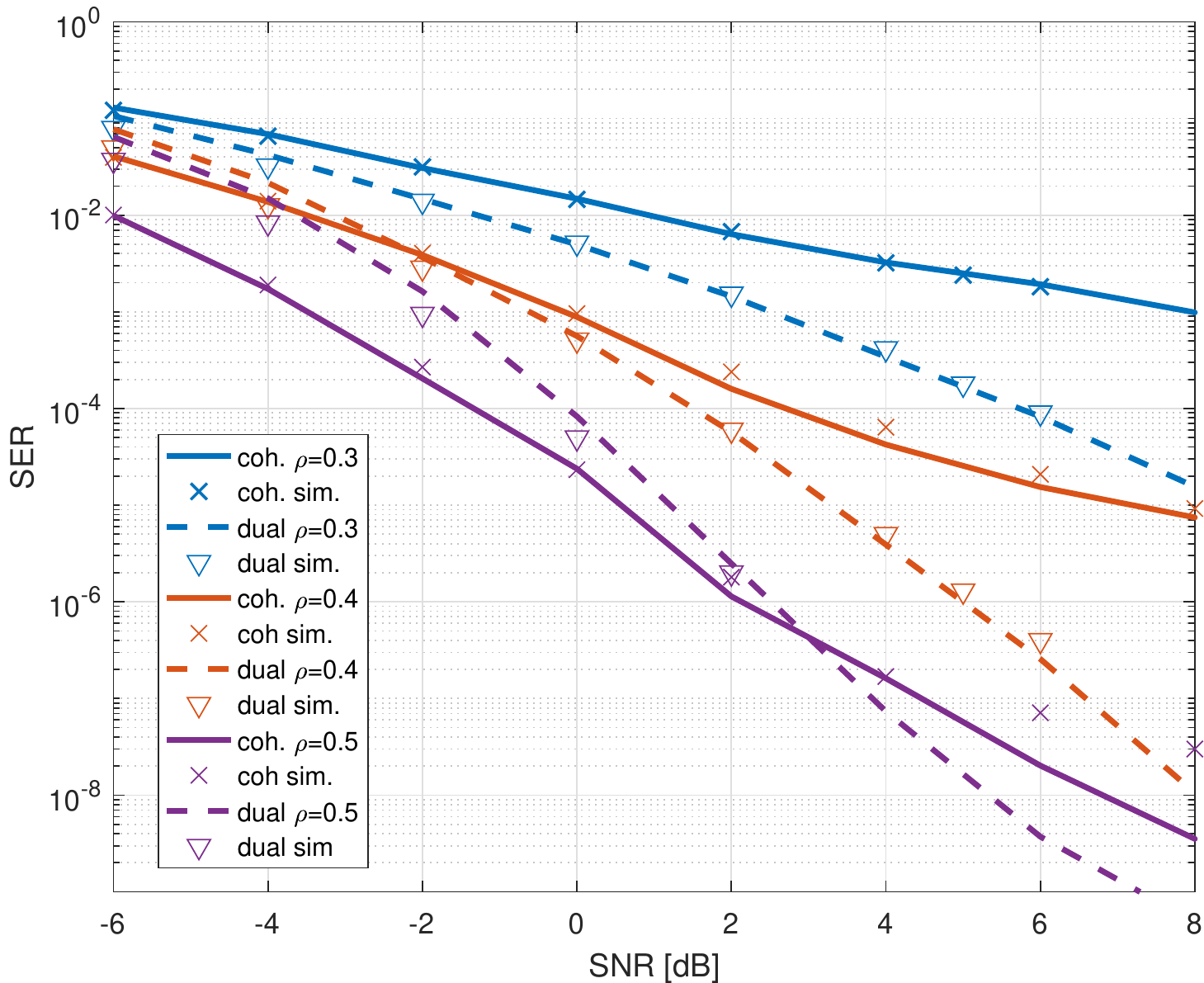}
    \caption{Comparison between coherent detection of 4-PSK (solid line) and dual-stage sequential non-coherent and coherent detection of 4-APK (dashed line) for constellation scaling $s=0.7$. Simulated points for the 4-PSK and dual-stage are denoted by crosses and triangles, respectively.}
    \label{fig:result1}
\end{figure}

The proposed dual-stage receiver with scaled constellation is compared to the 4-PSK coherent detector employing MRC, both through means of analysis and simulation, for several CSI correlation coefficients $\rho$. The result is shown in Fig.~\ref{fig:result1} for a constellation scaling coefficient $s=0.7$ and a number of 128 antennas at the receiver side. It can be observed that for inaccurate CSI, i.e. lower $\rho$, the dual-stage receiver provides considerable gain, especially towards higher SNRs, which can be the case of deployed URLLC devices which are subject to interference or suffer from outdated CSI. The improvements are due to the robustness of the non-coherent detection towards CSI impairments, and due to the choice of the constellation scaling.
The simulated points for the coherent receiver seem to be extremely close to the analytical result, which suggests that the Gaussian approximation of $y_\text{MRC}$ is justified. However, the analytical result for the dual-stage receiver seems to be slightly higher than the simulations, which is due to the approximation used in determining the SER of the first stage of energy detection. It can be noticed, however, that as the SNR increases, the offset between the analytical result and the simulation is decreased.

% The sequential receiver is compared to the coherent ML receiver for both scaled 4-APK constellation and 4-PSK by means of simulation, for several correlation coefficients of the estimated CSI and the channel coefficients. 
% The result is shown in Figure~\ref{fig:result1} for a constellation scaling coefficient $s=0.7$ and a number of 128 antennas at the receiver side. 
% It can be seen that due to scaling, the performance of the coherent ML receiver is decreased compared to the 4-PSK. 
% However, for inaccurate CSI, i.e. low $\rho$, it can be seen that the sequential receiver provides considerable gain. 
% This is due to the robustness of the energy detection, which is impaired less than the coherent detection by the CSI inaccuracy. 
% As the accuracy increases, it can be noticed that the sequential receiver reaches its limit, due to the fact that that the noise becomes the limiting factor for the power separation between the symbols.

\section{Discussion and conclusion}
This paper proposes a solution to improve symbol detection reliability in a scenario where the CSI is inaccurate. 
Inaccuracies can be caused by several factors such as high mobility, limited training resources, or pilot contamination from neighboring cells. 
The proposed solution is to use a 4-APK symbol constellation, where each two symmetric symbols reside on distinct power levels, such that a dual-stage non-coherent and coherent receiver can be employed.
The non-coherent I-ED stage of the detection reduces the symbol uncertainty by making use of the massive MIMO effect on the channel energy estimation and the noise hardening effect of the massive array without requiring precise CSI, therefore providing additional robustness when CSI imperfections are present. 

It is shown that the scaling of the constellation plays an important factor in both stages of the detection: the higher the scaling, the better performance of the non-coherent stage and the worse performance of the second stage. Therefore, after introducing this trade-off, the 4-PSK coherent detection was compared with the scaled 4-APK dual-stage receiver by means of simulation and analysis.
The dual-stage receiver offers lower symbol error rate towards higher SNR and lower CSI accuracy regime.

% Power separation between the symbols can be adjusted by varying $s$. The further apart the energy levels are, the more robust the ED stage will become. However, this will in turn decrease the robustness of coherently detecting the two coherent symbols lying on the lower energy level, as seen from Figure~\ref{fig:sys_model}. Therefore, there exists a trade-off between the reliability of first and second stages of the sequential detection procedure.
% Having highlighted the trade-off between the reliability of non-coherent and coherent stages of the detection, an optimal choice of the constellation scaling factor $s$ can be made, either depending on the CSI correlation $\rho$, or assuming $\rho$ varies slightly around a value which could be measured in practical deployments.
\bibliographystyle{IEEEtran}

\end{document}